\documentclass[a4paper,superscriptaddress,aps,prl,twocolumn,floatfix,citeautoscript,showpacs]{revtex4}
\usepackage{graphicx}
\usepackage{latexsym}
\usepackage{amsmath}
\usepackage{natbib}
\usepackage{mciteplus}
\usepackage{color}
\usepackage{float}

\newcommand{\vE}{\vec{E}}
\newcommand{\vA}{\vec{A}}
\newcommand{\vecr}{\vec{r}}
\newcommand{\vj}{\vec{j}}
\newcommand{\vP}{\vec{P}}

\begin{document}
\title{Fully parameter-free calculation of optical spectra for
insulators, semiconductors and metals from a simple polarization functional}

\newcommand{\lcpq}{Laboratoire de Chimie et Physique Quantiques, IRSAMC, Universit\'e Toulouse III - Paul Sabatier, 
CNRS and European Theoretical Spectroscopy Facility (ETSF), 118 Route de Narbonne, 31062 Toulouse Cedex, France}
\affiliation{\lcpq}

\author{J. A. Berger}
\email{arjan.berger@irsamc.ups-tlse.fr}
\affiliation{\lcpq}

\date{\today}

\pacs{71.15.Mb, 71.45Gm, 71.10-w, 71.15Qe}

\begin{abstract}
We present a fully parameter-free density-functional approach 
for the accurate description of optical absorption spectra of insulators, semiconductors and metals.
We show that this can be achieved within time-dependent current-density-functional theory using a simple dynamical polarization functional.
We derive this functional from physical principles that govern optical spectra.
Our method is truly predictive because not a single parameter is used.
In particular, we do not use an \textit{ad-hoc} material-dependent broadening parameter to compare theory to experiment as is usually done.
Our approach is numerically efficient; the cost equals that of a calculation within the random-phase approximation. 

\end{abstract}
\maketitle
%
The calculation of accurate optical absorption spectra within a density-functional approach has been
a major challenge for almost two decades.
While an accurate approach to calculate optical spectra is by solving the Bethe Salpeter equation (BSE) 
~\cite{Albrecht,Benedict,Rohlfing,OnidaReiningRubio}, the large numerical effort required precludes its application to large systems.
Therefore an approach based on the numerically more affordable time-dependent density-functional theory (TDDFT)~\cite{RungeGross}
or time-dependent current-density-functional theory (TDCDFT)~\cite{DharaGhosh,GhoshDhara,Vignale}
that gives optical spectra of similar quality as the BSE is much sought after.
The price to pay for the gained computational efficiency is that
the auxiliary Kohn-Sham system of TD(C)DFT is governed by an unknown effective potential
for which the derivation of accurate approximations is highly nontrivial.
The main failures of standard approximations 
such as the random-phase approximation (RPA)
and the adiabatic local-density approximation (ALDA)~\cite{ZangwillSoven} are:
1) the underestimation of the onset of the absorption; 
2) the underestimation of the intensity of continuum excitons; 
3) the absence of bound excitons;
4) the absence of Drude tails in the spectra of metals.
While the first problem can be circumvented by replacing the Kohn-Sham eigenvalues 
by quasiparticle energies obtained within the $GW$ approximation~\cite{HedinPR,AulburJonssonWilkins},
the other failures have proven to be more complicated to solve.
Early attempts to go beyond such standard approximations focused on a correct description of continuum excitons 
by employing a long-range exchange-correlation (xc) kernel either within TDCDFT~\cite{deBoeij} or TDDFT~\cite{RORO,Botti_2004}.
These approaches produced the desired result but at the cost of introducing a material-dependent parameter.
The nanoquanta kernel~\cite{Sottile,Marini_2003,Adragna,Stubner,vonBarth} 
does not require such a parameter and leads to a correct description of both
bound and continuum excitons. 
Unfortunately, being derived from the BSE, it is almost as expensive to evaluate as solving the full BSE.
An efficient and parameter-free TDCDFT approach can be obtained using the Vignale-Kohn functional~\cite{VK}.
While this method describes well the optical spectra of metals~\cite{Arjan2} 
it does not describe correctly excitons in semiconductors and insulators~\cite{Arjan1}.
Recently several new approaches have been proposed.
The bootstrap method advances an expression for the TDDFT xc kernel
which has to be obtained from a self-consistent procedure~\cite{bootstrap}.
In Ref.~[\onlinecite{Nazarov_2011}] an xc functional is derived from a meta-generalized-gradient approximation.
Finally, Ref.~[\onlinecite{Trevisanutto}] proposes an xc functional that is based on the jellium-with-gap model.
Although these new methods improve the description of excitonic effects, they have several shortcomings:
1) they are not parameter free due to an \textit{ad-hoc} material-dependent broadening parameter that is used to compare theory and experiment;
2) they are static and therefore do not account for memory effects precluding a parameter-free description of the finite width of Drude tails 
and bound excitons;
3) they require the calculation of the full Kohn-Sham density-density response function, an $O(N^4)$ calculation 
(with $N$ the number of atoms or electrons in the unit cell).

In this work we present an efficient $O(N^3)$ density functional approach that does not suffer from the shortcomings mentioned above. 
It is parameter free and dynamic, and it produces accurate optical spectra for metals, semiconductors, and insulators.
The method we propose is both simple and elegant and is characterized by the following two steps:
1) We calculate the macroscopic dielectric function using an efficient TDCDFT approach \cite{Freddie,Pina_2005}
within the RPA (including a scissors shift obtained from $GW$);
2) We apply a simple dynamical polarization functional (see Eq.~(\ref{Eqn:polfunc}) below) that accounts for continuum and bound excitons 
as well as Drude tails.
We will now give all the details of our approach. We use Hartree atomic units throughout.

The optical response of a solid can be obtained as the response to a homogeneous electric field, \textit{i.e.}, a field with vanishing wave vector $\vec{q}$.
However, for $\vec{q}=0$, the periodic density is not enough to uniquely describe the response; one also needs the macroscopic polarization~\cite{GGG_1995}.
This is an immediate consequence of the use of periodic boundary conditions.
Macroscopic effects due to the surface density have to be accounted for via the macroscopic polarization.
An elegant and efficient approach is to use the current density
as the fundamental quantity of a DFT for time-dependent phenomena.
The self-consistent perturbing potential is a unique functional of the periodic current density
since the latter contains both the periodic density of the bulk (via the continuity relation) and 
the macroscopic polarization, which is obtained as a bulk property using the definition,
%
\begin{equation}
\vP_{mac}(\omega) = \frac{-i}{\omega V } \int_{V } d\vecr \delta \vj (\vecr,\omega).
\label{Eqn:Pmac}
\end{equation}
Here $V$ is the volume of the unit cell.
The limit $\vec{q}\rightarrow 0$ can thus be performed explicitly, by setting $\vec{q}=0$, and hence knowledge of the current density suffices 
and no explicit calculation of Kohn-Sham response functions is needed.
The macroscopic polarization is induced by a macroscopic electric field $\vE_{mac}(\omega)$ which comprises 
both the externally applied field and the macroscopic induced electric field. 
The constant of proportionality is the electric susceptibility tensor $\tensor\chi_e(\omega)$ which is defined as
\begin{equation}
\vP_{mac}(\omega) = \tensor\chi_e(\omega) \cdot \vE_{mac}(\omega).
\label{Eqn:chie}
\end{equation}
The macroscopic dielectric tensor $\tensor\epsilon_M(\omega)$ can be obtained from the electric susceptibility according to 
\begin{equation}
\tensor\epsilon_M (\omega)= 1+4\pi \tensor\chi_e(\omega).
\label{Eqn:epsilon}
\end{equation}
Therefore, for a given $E_{mac}(\omega)$, knowledge of the induced current suffices to calculate $\tensor\epsilon_M (\omega)$
from Eqs.~(\ref{Eqn:Pmac})-(\ref{Eqn:epsilon}).

Within Kohn-Sham linear response the current density is given by
\begin{align}
\delta \vj(\vecr,\omega) &=
\notag
\int d\vecr' \tensor\chi^{s,\vj\vj}(\vecr,\vecr',\omega) [\vA_{mac}(\omega)+\vA_{mac}^{xc}(\vecr',\omega)] 
\\ &+
\int d\vecr' \vec\chi^{s,\vj\rho}(\vecr,\vecr'.\omega) \delta v^{Hxc}_{mic}(\vecr',\omega),
\label{Eqn:current}
\end{align}
%
where $\vA_{mac}(\omega)=\vE_{mac}(\omega)/(i\omega)$.
We note that the Kohn-Sham current-current and current-density response functions
($\chi^{s,\vj\vj}(\omega)$ and $\vec\chi^{s,\vj\rho}(\omega)$, respectively)
are never needed explicitly since we use a sum-over-states representation.
Therefore the $\vecr'$ integrals can be evaluated independently of $\vecr$.
In Eq.~(\ref{Eqn:current}) we used the microscopic Coulomb gauge
in which the microscopic potential $\delta v^{Hxc}_{mic}(\vecr,\omega)$ is lattice periodic and contains the periodic part of the
Hartree and longitudinal xc contributions ~\cite{Freddie}.
All remaining xc contributions are included in the the vector potential $\vA^{xc}(\vecr,\omega)$.
Here we neglect microscopic transverse xc contributions and therefore only its macroscopic part, $\vA_{mac}^{xc}(\omega)$, 
enters Eq.~(\ref{Eqn:current}).
It is related to the induced current through the TDCDFT tensor xc kernel $\tensor{f}_{xc}(\vecr,\vecr',\omega)$:
\begin{equation}
\vA_{mac}^{xc}(\omega) = \frac{1}{V}\int_{V} d\vecr \int d\vecr' \tensor{f}_{xc}(\vecr,\vecr',\omega)\cdot \delta\vj(\vecr',\omega).
\label{Eqn:fxc}
\end{equation}
Substitution of Eq.~(\ref{Eqn:current}) into Eq.~(\ref{Eqn:Pmac}), and using $\vE^{(xc)}_{mac}(\omega)=i\omega\vA^{(xc)}_{mac}(\omega)$, 
reveals that $\vP_{mac}(\omega)$ is linear in the macroscopic Kohn-Sham electric field,
$\vE_{mac}(\omega) + \vE_{mac}^{xc}(\omega)$.
Following Ref. [\onlinecite{deBoeij}], we define the auxiliary susceptibility $\tensor\chi_e^0$ according to
\begin{equation}
\vP_{mac}(\omega) = \tensor\chi_e^0(\omega) \cdot [\vE_{mac}(\omega) + \vE_{mac}^{xc}(\omega)].
\label{Eqn:chieKS}
\end{equation}
where $\tensor\chi_e^0$ is the susceptibility obtained from a calculation with $\vE_{mac}^{xc}(\omega)=0$.
We note that the contributions of $\delta v^{Hxc}_{mic}(\vecr,\omega)$ to $\tensor\chi_e^0$ are fully accounted for.
Since $\delta v^{Hxc}_{mic}(\vecr,\omega)$ is itself a functional of $\delta\vj(\vecr,\omega)$ this is done within a self-consistent field (SCF)
calculation. We will now show that the macroscopic xc effects can be accounted for post-SCF.
Comparison of Eqs.\ (\ref{Eqn:chie}) and (\ref{Eqn:chieKS}) leads to the following relation between $\tensor\chi_e$ and $\tensor\chi_e^0$,
\begin{equation}
\left([\tensor\chi_e^0]^{-1}(\omega) - [\tensor\chi_e]^{-1}(\omega)\right) \cdot \vP_{mac}(\omega) = \vE_{mac}^{xc}(\omega).
\label{Eqn:comparison}
\end{equation}
If one chooses $\vE_{mac}^{xc}(\omega)$ equal to zero as is usually done we obtain $\tensor\chi_e(\omega)=\tensor\chi_e^0(\omega)$.
We can now go beyond this approximation 
and obtain a polarization functional for $\vE_{mac}^{xc}(\omega)$ by neglecting microscopic current components 
in Eq.~(\ref{Eqn:fxc}), \textit{i.e.}, we replace $\delta\vj(\vecr,\omega)$ by its unit-cell average.
We obtain
\begin{equation}
\vE_{mac}^{xc}(\omega) = \tensor\alpha(\omega) \cdot \vP_{mac}(\omega)
\label{Eqn:Excmac}
\end{equation}
where we used Eq.~(\ref{Eqn:Pmac}) and in which we defined
\begin{equation}
\tensor\alpha(\omega) = -\frac{\omega^2}{V} \int_{V} d\vecr \int d\vecr' \tensor{f}_{xc}(\vecr,\vecr',\omega).
\end{equation}
Substitution of Eq.~(\ref{Eqn:Excmac}) into Eq.~(\ref{Eqn:comparison}) leads to
\begin{equation}
[\tensor\chi_e]^{-1}(\omega) = [\tensor\chi_e^0]^{-1}(\omega) - \tensor\alpha(\omega)
\label{Eqn:alpha}
\end{equation}
Therefore, for a given $\tensor\alpha(\omega)$, we can simply calculate $\tensor\chi_e(\omega)$ from $\tensor\chi_e^0(\omega)$.

Here we will use the RPA ($\delta v_{mic}^{Hxc} = \delta v_{mic}^{H}$) to calculate $\tensor\chi_e^0(\omega)$, \textit{i.e.}, $\tensor\chi_e^0=\tensor\chi_e^{RPA}$.
Since continuum excitons are underestimated and bound excitons and Drude tails are absent in RPA optical spectra,
we will include these effects through $\alpha(\omega)$.
We first derive a static $\alpha$ from the physical principles that govern strongly bound excitons in solids.
To simplify our argumentation we will assume for the moment that $\tensor\chi_e(\omega)$, $\tensor\chi_e^{RPA}(\omega)$ and
$\tensor\alpha(\omega)$ are isotropic, \textit{i.e.}, $\alpha_{ij}(\omega)=\alpha(\omega)\delta_{ij}$, etc..
Following arguments similar to those given in Ref.\ [\onlinecite{Sottile_2003}], we derive in the supplemental material~\cite{supmat}
that the exact constraint to have a bound exciton is that for some $\omega=\omega_{be}$ below the band gap the following
relation holds for $\alpha(\omega)$,
\begin{equation}
\textrm{Re}[\alpha(\omega_{be})] = 1/\chi^{RPA}_e(\omega_{be}) \quad \mathrm{and} \quad \mathrm{Im}[\alpha(\omega_{be})]\simeq 0.
\label{Eqn:alphastaticexact}
\end{equation}
However, the application of the above expression requires the knowledge of $\omega_{be}$ which is unknown.
Therefore, we would like to rewrite the above expression in terms of known quantities.
In the supplemental material~\cite{supmat} we derive such an expression for the case 
of a wide-gap insulator whose optical spectra is dominated by a bound exciton.
The result is a static $\alpha$ given by
%
\begin{equation}
\alpha =
\frac{1}{\chi^{RPA}_e(\omega=0)\epsilon_M^{RPA}(\omega=0)}.
\label{Eqn:alphastatic}
\end{equation}
Although the derivation applies to wide-gap insulators, 
Eq.~(\ref{Eqn:alphastatic}) has the additional advantage that $\alpha$ is proportional to $[\epsilon_M^{RPA}(0)]^{-1}$.
It has previously been shown numerically that such a proportionality leads to good optical spectra for semiconductors
~\cite{Botti_2004,Botti_2005}.
The physical reason is that $\epsilon_M^{RPA}(0)$ is a measure of the amount of screening of the electron-hole interaction.
Screening effects are more important in semiconductors than in wide-gap insulators.
We note that Eq.~(\ref{Eqn:alphastatic}) has a similar form as the bootstrap kernel~\cite{bootstrap} 
but that no self-consistent procedure is needed to calculate it.
Unfortunately $\alpha$ in Eq.~(\ref{Eqn:alphastatic}) is static and will therefore not be able to account 
for the finite width of bound excitons and Drude tails.
For this reason we add to Eq.~(\ref{Eqn:alphastatic}) $Y_{VK}(\omega)$, the long-range part of the dynamical Vignale-Kohn functional~\cite{VK} 
in which we replace the current density by its unit-cell average~\cite{deBoeij}:
\begin{align}
& \tensor{Y}_{VK}(\omega) = \frac{1}{V} \int_{V} d\vecr
\Bigg(
\frac{\nabla\rho_0(\vecr)\cdot\nabla\rho_0(\vecr)}{\rho_0^2(\vecr)} f_{xcT}(\bar\rho,\omega) \tensor{\mathrm{I}}
\notag \\ & + 
\frac{\nabla\rho_0(\vecr)\otimes\nabla\rho_0(\vecr)}{\rho^2_0(\vecr)} 
\left[f_{xcL}(\bar\rho,\omega)-f_{xcT}(\bar\rho,\omega)-\frac{d^2 e_{xc}}{d\bar\rho^2}\right]
\Bigg).
\label{Eqn:YVK}
\end{align}
Here $f_{xcL(T)}(\omega)$ is the longitudinal (transverse) xc kernel of the homogeneous electron gas, $e_{xc}$ is the xc energy per volume 
of the homogeneous electron gas, $\rho_0(\vecr)$ is the ground-state density and $\bar\rho$ is its average in the unit cell.
For $f_{xcL(T)}(\omega)$ we use the parametrization of Ref.\ [\onlinecite{QianVignale}] in the QVA approximation~\cite{Arjan2}.
The Vignale-Kohn functional is the exact functional of a slightly inhomogeneous electron gas and describes correctly 
the optical spectra of metals~\cite{Arjan2}.
For this reason Eq.~(\ref{Eqn:YVK}) is complementary to Eq.~(\ref{Eqn:alphastatic}) which tends to zero for metallic systems
because screening is complete and therefore $[\epsilon^{RPA}_M(\omega=0)]^{-1}\rightarrow 0$.
$\tensor{Y}_{VK}(\omega)$ will account for the Drude tails and finite width of bound excitons which are 
exactly the features that are absent in Eq.~(\ref{Eqn:alphastatic}). 
Moreover, as we show in the supplemental material~\cite{supmat},
$\tensor{Y}_{VK}(\omega)$ will not much influence the spectra of semiconductors.
We finally obtain the following approximation for $\tensor\alpha(\omega)$,
\begin{equation}
\tensor\alpha(\omega) = [\tensor\epsilon_M^{RPA}(0)]^{-1} [\tensor\chi_{e}^{RPA}(0)]^{-1} + \tensor{Y}_{VK}(\omega)
\label{Eqn:polfunc}
\end{equation}
where we generalized Eq.~(\ref{Eqn:alphastatic}) to a tensor form.
Since $[\tensor\epsilon_M^{RPA}(0)]^{-1}$ and $[\tensor\chi_{e}^{RPA}(0)]^{-1}$ commute,
the order of the multiplication is irrelevant.
We note that Eq.~(\ref{Eqn:polfunc}) satisfies the Kramers-Kronig relations.
Equation (\ref{Eqn:polfunc}) is the main result of this work.

We will now demonstrate our approach by applying it to the calculation of the optical spectra of several materials.
We briefly outline the full procedure.
The ground-state calculations are done within the local-density approximation (LDA)~\cite{KohnSham}
and we use LDA lattice parameters in all calculations.
We apply a scissors operator to shift the unoccupied bands and we modify the current operator accordingly
to guarantee that exact constraints such as the continuity equation remain satisfied.
The energy shift is calculated with the $GW$ method~\cite{HedinPR,AulburJonssonWilkins} and is equal to the $G_0W_0$ correction 
for the direct band gap at the $\Gamma$ point.
We calculated this correction using the effective-energy technique~\cite{EET1,EET2,EET3},
which leads to a speed-up of an order of magnitude.
The calculations were done with the Abinit code~\cite{Abinit}.
We implemented the polarization functional in a modified version of the Amsterdam Density Functional (ADF) code ~\cite{ADF1,ADF2,ADF3}.
We use the TZ2P (triple-$\zeta$ + 2 polarization functions) basis set provided by ADF.
The $\vec{k}$-space integrals are done analytically using a Lehmann-Taut tetrahedron scheme~\cite{LehmanTaut}.
Since we do not include effects due to electron-phonon coupling the spectra obtained with the above approach are predictions of 
the optical spectra at low temperature where electron-electron scattering dominates electron-phonon scattering.
For this reason we will compare our calculated spectra with spectra measured at low temperature where available.
\begin{figure}[t]
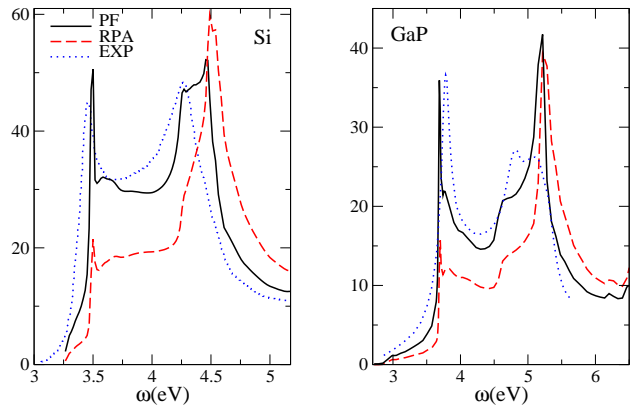

\centering
\rotatebox{90}{\parbox{0.1in}{$\epsilon_2(\omega)$}}
\parbox{1.2in}{
\includegraphics[width=0.43\columnwidth]{Si.eps}
}
\qquad\qquad
\parbox{1.2in}{
\includegraphics[width=0.43\columnwidth]{GaP.eps}
}
\caption{(Color online) The optical absorption spectra of bulk silicon and GaP. 
Solid line (black): polarization functional (PF); Dashed line (red): RPA; 
Dotted line (blue): experiment from Ref.\ [\onlinecite{Lautenschlager_sil}] (Si) and Ref.\ [\onlinecite{Zollner_gap}] (GaP).}
\label{Fig:Siandgap}
\end{figure}

In Fig.~{\ref{Fig:Siandgap}} we report the optical absorption spectra of bulk silicon and bulk GaP obtained with our polarization functional 
and compare it to the RPA spectra and to experimental results obtained at low temperature (15 Kelvin).
Silicon and GaP are typical examples of materials for which the RPA strongly underestimates the first peak
which appears in the experimental spectra around 3.4 eV (Si) and 3.8 eV (GaP).
Our polarization functional solves this problem by including the necessary excitonic effects and the first peak compares
well with experiment both in position and magnitude. Overall, the spectra are very close to experiment with the exception
of the peak around 5.2 eV in the spectrum of GaP which is overestimated.

\begin{figure}[t]
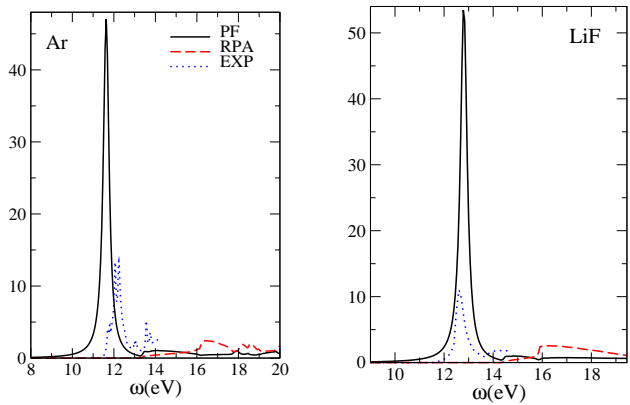

\centering
\rotatebox{90}{\parbox{0.1in}{$\epsilon_2(\omega)$}}
\parbox{1.2in}{
\includegraphics[width=0.43\columnwidth]{Ar.eps}
}
\qquad\qquad
\parbox{1.2in}{
\includegraphics[width=0.43\columnwidth]{LiF.eps}
}
\caption{The optical absorption spectra of solid Argon and LiF.
Solid line (black): polarization functional (PF); Dashed line (red): RPA; 
Dotted line (blue): experiment from Ref.\ [\onlinecite{Saile_Ar}] (Ar) and Ref.\ [\onlinecite{Roessler_lif}] (LiF).}
\label{Fig:Arandlif}
\end{figure}

In Fig.~{\ref{Fig:Arandlif}} we show the optical absorption spectra of solid argon and LiF obtained with our polarization functional and compare
it to the RPA spectra and to experimental results.
Solid argon and LiF are typical materials that exhibit strongly bound excitons.
We see that these excitons which appear in the experimental spectra around 12 eV (Ar) and 12.5 eV (LiF) are completely absent in the RPA spectra.
Our polarization functional describes these bound excitons and also accurately reproduces their position.
The magnitude of the peaks is overestimated with respect to experiment.
This discrepancy is probably due to electron-phonon broadening in the experiments and
to the fact that density-functional approaches tend to overestimate these peaks ~\cite{bootstrap,Trevisanutto}.
Finally, we verify \textit{a posteriori} that $\alpha$ given in Eq.~(\ref{Eqn:alphastatic})
is indeed a good approximation to Eq.~(\ref{Eqn:alphastaticexact}) for wide-gap insulators
exhibiting a dominant bound exciton. With Eq.~(\ref{Eqn:alphastatic}) we obtain 
$\alpha_{\mathrm{LiF}}=8.9$ and $\alpha_{\mathrm{Ar}}=11.8$, while Eq.~(\ref{Eqn:alphastaticexact}) gives 
$\alpha_{\mathrm{LiF}}=9.2$ and $\alpha_{\mathrm{Ar}}=12.2$.

\begin{figure}[t]
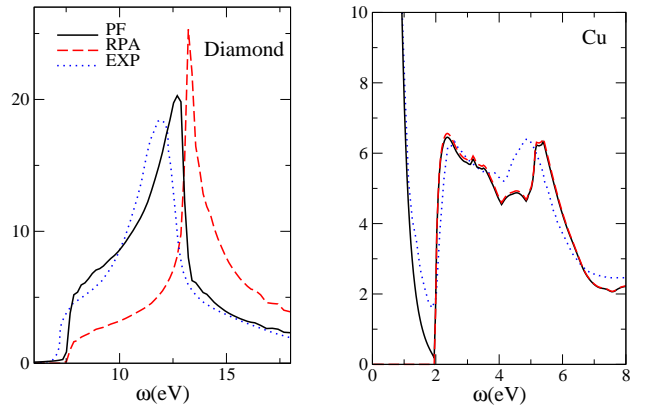

\centering
\rotatebox{90}{\parbox{0.1in}{$\epsilon_2(\omega)$}}
\parbox{1.2in}{
\includegraphics[width=0.43\columnwidth]{C.eps}
}
\qquad\qquad
\parbox{1.2in}{
\includegraphics[width=0.43\columnwidth]{Cu.eps}
}
\caption{The optical absorption spectra of diamond and copper.
Solid line (black): polarization functional (PF); Dashed line (red): RPA; 
Dotted line (blue): experiment from Ref.\ [\onlinecite{Philip_C}] (Diamond) 
and Refs.\ [\onlinecite{Stahrenberg_cu}] and [\onlinecite{Hagemann_cu}] (Cu).}
\label{Fig:Candcu}
\end{figure}

In Fig.~{\ref{Fig:Candcu}} we report the optical absorption spectra of diamond and copper obtained with the polarization functional 
and compare it to the RPA spectra and to experimental results.
Diamond is another typical test case since the RPA spectrum is quite different from the experimental spectrum.
Due to the absence of excitonic effects the RPA spectrum has too much weight at high energy.
With our polarization functional the spectral weight is shifted to lower energy and we obtain a very good agreement with experiment.
While the RPA spectrum of copper accurately reproduces the part of the spectrum which is due to interband transitions, the Drude tail at
low energy, which is due to intraband transitions, is completely absent. 
Our polarization functional accurately describes the Drude tail while maintaining the good agreement for the interband part.

In conclusion, we presented the first fully parameter-free density-functional approach that gives accurate optical spectra
for insulators, semiconductors and metals alike.
Our approach is therefore truly predictive and due to its numerical efficiency opens the way for 
the prediction of optical spectra of large systems.

We thank Paul L.\ de Boeij, Lucia Reining, and Pina Romaniello for fruitful discussions.
\bibliography{/home/berger/Documents/Tex/Articles/kernel/Resub/Article.bib}
\end{document}